\documentclass[aps,showpacs]{revtex4}
\usepackage{epsf,psfig}
\def\nl{\nonumber \\}
\def\gev{{\mathrm{GeV}}}

\def\order#1{{\cal O}\!\left(#1\right)}
\newcommand{\ba}{\begin{eqnarray}}
\newcommand{\ea}{\end{eqnarray}}
\newcommand{\be}{\begin{equation}}
\newcommand{\ee}{\end{equation}}

\def\eq#1{(\ref{#1})}
\def\eqn#1{eq.~(\ref{eq#1})}
\def\vud{\left|V_{ud}\right|}
\def\vus{\left|V_{us}\right|}
\def\vub{\left|V_{ub}\right|}
\def\tn{\tau_n}
\def\ke3{$K_{e3}$}
\def\g{\hat{g}}

\begin{document}

\title{
%
%
\[ \vspace{-2cm} \]
\noindent\hfill\hbox{\rm \normalsize Alberta Thy 12-04}
\vskip 1pt
\noindent\hfill\hbox{\rm \normalsize BNL-HET-04/7}
\vskip 1pt
\noindent\hfill\hbox{\rm \normalsize NYU-TH/04-06-25}
\vskip 10pt
Precision Measurements and  CKM Unitarity}

\author{Andrzej Czarnecki}
\affiliation{
Department of Physics, University of Alberta\\
Edmonton, AB,   Canada T6G 2J1}

\author{William J. Marciano}
\affiliation{Brookhaven National Laboratory,
Upton, NY 11973, USA }

\affiliation{Institut f\"{u}r Theoretische Teilchenphysik,
Universit\"{a}t Karlsruhe,
D--76128 Karlsruhe, Germany}

\author{Alberto Sirlin}
\affiliation{Department of Physics, New York University,\\
    4 Washington Place, New York, NY 100003, USA}

\affiliation{{II. I}nstitut f\"ur Theoretische Physik, Universit\"at Hamburg,
\\    Luruper Chaussee 149, 22761 Hamburg, Germany}

\begin{abstract}
Determinations of $\vud$ and $\vus$ along with their implications for the
unitarity test $\vud^2+\vus^2+\vub^2=1$ are  discussed.
The leading two loop radiative corrections to neutron $\beta$-decay are
evaluated and used to derive a refined relationship  $\vud^2
\tau_n (1+3g_A^2)=4908(4)$sec.  Employing $\vud =0.9740(5)$
from superallowed nuclear decays and the measured neutron
lifetime $\tau_n = 885.7(7)$sec, leads to the precise prediction
$g_A=1.2703(8)$ which is compared with current direct experimental
values.  Various extractions of $\vus$ are described and updated.
The long accepted
Particle Data Group value of $\vus$
from fitted $K_{e3}$ decay rates suggests a
deviation from CKM unitarity but it is
contradicted by more recent experimental
results which confirm unitarity with good precision.
 An outlook for possible future advances is given.
\end{abstract}


\maketitle

\section{Introduction}

The study of nuclear beta decays played an important historical
role in unveiling universal properties of weak charged current
interactions and in helping to establish the $SU(2)_L\times
U(1)_Y$ Standard Model of electroweak unification.  In the limit of neglecting
electroweak loop corrections, a  special
subset of such decays, the superallowed $0^+ \to 0^+$ Fermi
transitions, depend only on the vector current which is conserved
and, therefore, not renormalized by strong interactions at $q^2
\simeq 0$.  Hence, they are ideal for extracting $\vud$, a
cornerstone of the CKM (Cabibbo-Kobayashi-Maskawa \cite{Cabibbo:1963yz,Kobayashi:1973fv})
three generation quark mixing matrix,
\ba
V_{\rm CKM} =
\left(
\begin{array}{lll}
V_{ud} & V_{us} & V_{ub} \\
V_{cd} & V_{cs} & V_{cb} \\
V_{td} & V_{ts} & V_{tb}
\end{array}
\right).
\label{eq1}
\ea
Indeed, those decays currently provide the very precise determination \cite{Towner:2003cq}
\ba
\vud = 0.9740(5) \qquad (0^+\to 0^+ \ \beta\mbox{-decays}),
\label{eq2}
\ea
which we will discuss later in this paper.
Combining the value in eq.~\eq{eq2} with knowledge about $\vus$
from kaon and Hyperon decays along with the fact that $\vub^2\simeq
2.1(3)\cdot 10^{-5}$ \cite{Aubert:2003zw}
is negligibly small, allows one to confront
the CKM unitarity relation,
\ba
\vud^2+\vus^2+\vub^2=1,
\label{eq3}
\ea
at a high precision level.

That prediction has been tested to about $\pm 0.15\%$, an
impressive accomplishment. At that level, it has
confirmed the presence of very large, $\sim 4\%$,
Standard Model loop corrections
in the extraction of $\vud^2$ from the data  \cite{Sirlin:1974ni,Sirlin:1978sv}.
However, for many years, a small deviation, $\sim
2\sigma$, from exact unitarity has been persistently observed.
That issue has been further clouded by various distinct $\vus$
determinations which seem to be inconsistent with one another.
 These problems are addressed later in some detail,
when new \ke3 results fully consistent with unitarity are also discussed.
Of course, if a real deviation from unitarity expectations is
seen, it signals the presence of as yet unaccounted for new
physics beyond the Standard Model, an exciting prospect.  Alternatively,
if unitarity is respected, constraints on new physics can
be implied.  However,
$\vud$ and $\vus$ must be thoroughly scrutinized both theoretically
and experimentally before conclusions are drawn.

Neutron beta decay, $n\to p e^- \overline \nu_e$, has not until
recently been prominent in efforts to determine $\vud$  and test
unitarity.  It depends on both vector and axial-vector charged
current interactions.  The latter are renormalized by strong
interactions at $q^2\simeq 0$. The size of that effect is
parameterized by $g_A\equiv G_A/G_V$, a fundamental quantity in
its own right.  Indeed, the value of $g_A$, which has grown over
time from about 1.23 to 1.27, is important for predicting the
expected solar neutrino flux \cite{Adelberger:1998qm},
light element abundances from
primordial nucleosynthesis \cite{Esposito:1998rc},
the spin content of nucleons \cite{Close:1988de,Close:1993mv} and for
testing the Goldberger-Treiman relation \cite{Goldberger:1958tr}.  A byproduct of the
analysis in this paper will be to provide a very precise
determination of $g_A$ that can be compared with more direct
neutron decay asymmetry measurements of that important parameter
or to refine the above mentioned applications.

As new more intense neutron facilities turn on, experimental
measurements of both the neutron lifetime $\tn$ and $g_A$ from
the electron asymmetry in polarized neutron beta decay are
expected to become much more precise \cite{Abele:2003pz}.  Indeed, combining
determinations of those two quantities, can yield $\vud$  with an
anticipated uncertainty competitive with the error in
eq.~\eq{eq2}, i.e. dominated by theory.  In preparation for those
improvements, we present in this paper a relationship among
$\vud$, $\tn$ and $g_A$ which includes one and some dominant two loop
quantum corrections. It can be used to determine $g_A$ from $\tn$
and the $\vud$ input from superallowed nuclear $\beta$-decays or
as an independent measure of $\vud$ using $\tn$ and neutron decay
asymmetry determinations of $g_A$.

Our plan is as follows: in section \ref{sec2}, we update the
radiative corrections to neutron decay by incorporating the
$\order{\alpha^2}$ effects due to leading logs, (some) small
next-to-leading logs and Coulombic effects.  The last of these has
been considered previously, but with the wrong sign.  We take this
opportunity to correct that longstanding error. In section
\ref{sec3}, we review and update (slightly) the extraction of
$\vud$ from superallowed $\beta$-decays.  Using that value of
$\vud$ along with the neutron lifetime $\tn$ we derive in
section \ref{sec4} a very precise prediction for $g_A$ and compare
it with direct asymmetry measurements of that parameter.  Then, in
section \ref{sec5}, we review and update various determinations of
$\vus$ and point out inconsistencies among them.
The main problem stems from old \ke3 decay rates obtained from fitted PDG studies
and is suggestive of errors in some longstanding (accepted) kaon decay properties.
Indeed, recent \ke3 results from Brookhaven and Fermilab experiments confirm
significant errors in the old charged and neutral kaon decay branching ratios and lead to
values of $\vus$ fully consistent with unitarity.  Implications of
a unitarity violation or confirmation in \eqn{3} are briefly discussed and an
outlook for future advances is given
in section \ref{sec6}.

\section{Radiative Corrections to Neutron Decay}
\label{sec2} Our analysis of the radiative corrections to neutron
beta decay builds on the results of earlier studies, particularly
the classic work by Wilkinson \cite{Wilkinson:1982hu}.  They
included $\order{\alpha}$ radiative corrections as well as effects
due to the final state electromagnetic $ep$ interaction embodied
in the Fermi function. A number of other small corrections from
proton recoil, finite nucleon size etc. have  also been examined
\cite{Garcia:2000gk,Bernard:2004cm,Ando:2004rk}.

In the Standard Model, one renormalizes the beta decay amplitude
using
\cite{Kinoshita:1959ru,Berman:1958ti,vanRitbergen:1998yd,%
Steinhauser:1999bx,Ferroglia:1999tg,Blokland:2004ye}
\ba
G_\mu = 1.16637(1) \cdot 10^{-5} \ \mbox{GeV}^{-2},
\label{eq4}
\ea
the Fermi constant as obtained from the muon total decay rate
(inverse lifetime).  In that way, ultraviolet divergences as well
as radiative corrections common to both decay amplitudes are
absorbed into $G_\mu$.  The remaining loop differences and
bremsstrahlung effects can then be factorized into an overall 1+RC
correction to the neutron lifetime, $\tn$
\ba
{1\over \tn} = {G_\mu^2\vud^2\over 2\pi^3} m_e^5 \left(1+3g_A^2\right)
\left( 1+\mbox{RC}\right) f,
\label{eq5}
\ea
where $f$ is a phase space factor,
\ba
f=1.6887,
\label{eq6}
\ea
which includes a relatively large Fermi function contribution
\cite{Wilkinson:1982hu}
$(\sim 5.6\%)$ as well as smaller nucleon mass, size and recoil
corrections.  It has been somewhat updated in eq.~\eq{eq6} to
incorporate slight nucleon mass shifts.

We note that the electroweak radiative corrections, denoted by
1+RC, have been factorized in the same way for both the vector and
axial-vector contributions.  (Interference and induced coupling
corrections are negligibly small in the case of the lifetime
\cite{Wilkinson:1982hu,Garcia:2000gk}.)  That factorization effectively
defines $g_A$ via the relative  normalization of the
axial-vector current as measured by the lifetime (it incorporates
QED as well as strong interaction effects in its definition).
Employing such a definition for $g_A$ means that there will be
some $\order{\alpha}$ corrections to the $g_A^{\mathrm{asy}}$
measured in neutron decay asymmetries that must be applied before
contact with eq.~\eq{eq5} can be made at a level of high
precision. (Those corrections will be discussed in section \ref{sec4}.)
 In this connection, it is worthwhile to note that the normalization of $g_A$
in eq.~\eq{eq5} is consistent with the so-called $1/k$ method, in which the radiative
corrections to various observables in $\beta$-decay, such as the lifetime, the electron
spectrum, and the longitudinal electron polarization are expressed in terms of effective couplings
$G_V$ and $G_A$ \cite{Sirlin67,SirlinAustria,Blin-Stoyle:1970sz,Sirlin:2003ds}.
The same approach has been employed to calculate the corrections to the electron
asymmetry \cite{Shann,Yokoo,Garcia:1978bq,Gluck:1992qy,Fukugita:2004pq}.
The factorization of the short distance contributions of the radiative
corrections, implicit in \eqn{5}, conforms also with an asymptotic theorem
concerning their behavior in arbitrary semileptonic decays mediated by
$W^\pm$ \cite{Sirlin82}.

To order $\alpha$, the radiative corrections in eq.~\eq{eq5} are given by
\cite{Sirlin:1978sv,Marciano:1986pd,Sirlin:1974ni,Marciano:1993sh}
\ba
{\mathrm{RC}} = {\alpha\over 2\pi} \left\{
\overline g(E_m) + 4\ln{m_Z\over m_p} + \ln{m_p\over m_A} +2C +A_g\right\},
\label{eq7}
\ea
where $\overline g(E_m)$ represents long distance loop corrections and bremsstrahlung effects
averaged over the $\beta$-decay spectrum \cite{Sirlin67}
($E_m=1.292579$ MeV, the end-point electron energy in neutron decay),
\ba
{\alpha\over 2\pi}
\overline g(E_m)=0.015056.
\label{eq8}
\ea
The other parameters in that expression have the values
\ba
m_Z&=& 91.1875 \ \gev, \nonumber \\
m_p &=& 0.9383 \ \gev ,\nl
m_A &\simeq & 1.2 \ \gev ,\nl
C &\simeq & 0.891, \nl
A_g &\simeq & -0.34.
\label{eq9}
\ea
Here, following \cite{Towner:2003cq,Sirlin:2003ds,SirlinInLang}, we have approximately identified
$m_A$ with the mass of the $a_1(1260)$ axial vector meson.  The value of $C$ is an update of the
calculation discussed in Ref. \cite{Marciano:1986pd} (using $g_A = 1.27$ for self consistency).
 This leads to a first order result
\ba
{\rm RC}\left(\order{\alpha}\right) = 0.03770,
\label{eq10}
\ea
which is rather sizeable.  (Together with the Fermi function, such
(primarily) QED corrections increase the neutron decay rate by
9.37\%!)

Given the magnitude of the  order $\alpha$ corrections
in eq.~\eq{eq10} and our desire for high precision, it becomes
imperative to include the leading $\order{\alpha^2}$
contributions and estimate the theoretical uncertainty in the
radiative corrections.  Regarding the latter issue, the overall
uncertainty is usually obtained by allowing $m_A$ in eq.~\eq{eq7}
to vary by a factor of 2 up or down.  That reflects the fact that
the last 3 terms in eq.~\eq{eq7} result from axial-vector
current loop effects and their calculation is not perfectly
matched in going from long to short-distance contributions.
The scale, $m_A$, uncertainty reflects the matching error in
a rough, but numerically realistic, way.
With that methodology, the theoretical uncertainty
is estimated to be
\ba
\mbox{RC (uncertainty)} = \pm 0.0008,
\label{eq9b}
\ea
an error common to all neutron and nuclear
$\beta$-decay studies.  Reducing that
dominant theory matching error would be very useful, but it
is extremely challenging and beyond
the scope of this paper.

Our focus in this section is to include the so called leading log
corrections of the form $\alpha^n \ln^n{m_Z\over m_p}$ and
$\alpha^n \ln^n{m_p\over 2E_m}$, $n=2,3,\ldots$ in eq.~\eq{eq7}
along with some of the other potentially most important $\order{
\alpha^2 \ln{m_Z\over m_p}}$  and $\order{\alpha^2}$ effects.
Regarding the last of those, there is a relatively important
Coulomb correction not included in the product of the Fermi function
and 1+RC.  It is called ${\alpha\over 2\pi}\delta$ in the
literature and approximated by (see Appendix)
\ba
{\alpha\over 2\pi}\delta \simeq -\alpha^2 \ln {m_p\over m_e}
+\ldots = -0.00043.
\label{eq10b}
\ea
Although considered previously, for some reason it was given the
wrong sign.  Correcting the value $+0.0004$ used in the past
to eq.~\eq{eq10b} corresponds to a  $-0.00083$
shift which is quite significant at the level of our analysis.

In the Appendix,
we give formulas that sum the leading log
contributions in eq.~\eq{eq7} via the method in ref.~\cite{Marciano:1986pd}.
They lead to the replacements
\ba
1+{2\alpha\over \pi}\ln{m_Z\over m_p}& \to & S(m_p,m_Z) = 1.02248,
\nl
1+{3\alpha \over 2\pi}\ln {m_p\over 2E_m}&\to & L(2E_m,m_p)
=1.02094.
\label{eq11}
\ea
where the large ${3\alpha \over 2\pi}\ln {m_p\over 2E_m}$
contribution is hidden in the ${\alpha\over 2\pi}\overline
g(E_m)$ function of \eqn{7} \cite{Sirlin67}.

Our final new input is to estimate next-to-leading log (NLL) corrections
of the form $\alpha^2 \ln{m_Z\over m_p}$ and $\alpha^2 \ln
{m_p\over  m_f}$ coming from fermion vacuum polarization
insertions in loops with photon propagators.  Because they all
enter with the same sign and there are quite a few leptons
and quarks that contribute, one might expect those fermion
loops to dominate the NLL contribution.  However, as illustrated
in the Appendix,
they turn out to be quite small in the $\overline{\rm
MS}$ formalism we employ.  We estimate
\ba
{\rm NLL}=-0.0001.
\label{eq12}
\ea
Other $\order{\alpha^2}$ contributions are not expected to be
significant, but a complete calculation to $\order{\alpha^2}$
would be very difficult and beyond the scope of this paper.
Also, such a refined calculation is not obviously warranted until the one loop
matching uncertainty in \eqn{9b} is significantly reduced.

Our next step is to organize the radiative corrections into a
factorized form that does not induce spurious contributions when
multiplied out.  To accomplish that end, we take an effective
field theory approach, dividing contributions into very long
distance (the Fermi function), long distance, intermediate
distance, and short distance factors.  Of course, we must make
certain that the matching is done correctly, for example by
introducing terms such as ${\alpha\over 2\pi} \delta$ (see
eq.~\eq{eq10b}) when appropriate.  Overall, we employ the following
factorization beyond the Fermi function
\ba
1+{\mathrm{RC}} &=& \left[ 1+{\alpha\over 2\pi} \left( \overline g(E_m)
-3\ln{m_p\over 2E_m} \right)\right]
\nl
&& \cdot \left[ L(2E_m,m_p)+{\alpha\over \pi}C + {\alpha\over 2\pi} \delta\right]
\nl
&& \cdot \left[ S(m_p,m_Z) + {\alpha(m_p)\over 2\pi}\left(
\ln {m_p\over m_A} +A_g\right)+{\rm NLL}\right],
\label{eq13}
\ea
where $\alpha(m_p)\simeq 1/134$.  Employing the values we have given above for the
quantities in eq.~\eq{eq13}, one finds
\ba
1+{\mathrm{RC}} &=& 1.0390(8),
\label{eq14}
\ea
where the uncertainty in eq.~\eq{eq9b} has been employed.
Comparing eqs.~\eq{eq14} and \eq{eq10}, one sees that the
$\order{\alpha^2}$ and summation effects have increased the RC by
0.0013, not a very large shift.  Nevertheless, they must be
included in precision studies.
In summary, our updated analysis includes the following two-loop, $\order{\alpha^2}$
contributions: 1) Leading Logs, 2) Next-To-Leading Logs from fermion
vacuum polarization effects and 3) Coulombic matching corrections due to
the factorization of the Fermi function. Other neglected two-loop
corrections are not enhanced by relatively large factors and thus
assumed to be negligible.

Employing eq.~\eq{eq14} in eq.~\eq{eq5}, we derive the relationship
\ba
\vud^2 \left( 1+3g_A^2\right) \tn = 4908\pm 4\ \mbox{sec}.
\label{eq15}
\ea
That master formula can be used to extract $\vud$ via
\ba
\vud =\left( {4908(4)\ \mbox{sec}\over  \tn \left( 1+3g_A^2\right)}\right)^{1/2},
\label{eq16}
\ea
as $\tn$ and $g_A$ become more precise or to obtain $g_A$ from $\tn$ and $\vud$.  Currently, one
finds from the experimental averages
\ba
\tn^{\rm ave} &=& 885.7(7)\ \mbox{sec}
\label{eq17a}
\\
g_A^{\rm ave} &=& 1.2720(18),
\label{eq17b}
\ea
the CKM parameter
\ba
\vud = 0.9729(4)(11)(4),
\label{eq18}
\ea
where the errors correspond to $\tn$, $g_A$ and RC respectively.
In section \ref{sec3}, we compare that value with the more
precise $\vud$ obtained from superallowed
beta decays. Of course, the full utility of eq.~\eq{eq16} will
be much better realized as $\tn$ and $g_A$ measurements improve.

The master formula in eq.~\eq{eq15} can also be used to extract a very precise value of $g_A$
via
\ba
g_A = \left[ {1636(1) \ \mbox{sec}\over \tn \vud^2} - {1\over 3}\right]^{1/2}.
\label{eq19}
\ea
Note that the theory
uncertainty ($\pm 1$ sec)
in \eqn{19}   is due to the $m_A$ scale uncertainty in
the radiative corrections and it cancels with the same error in
determinations of $\vud^2$ from nuclear decays.
The formula in eq.~\eq{eq19} will be utilized in section \ref{sec4}.

\section{Superallowed $\beta$-decays and $\vud$}
\label{sec3}

Superallowed $0^+\to 0^+$ Fermi transitions have been the focus of
many studies, both experimentally and theoretically.  Here, we
start with the recent results of Towner and Hardy \cite{Towner:2002rg}.  They have
thoroughly scrutinized the nine very well measured superallowed
$\beta$-decays,
using RC last updated in ref. \cite{SirlinInLang,Sirlin:1987sy,Jaus:1987te}, and
 taking great care to correct for various nuclear
Coulombic and structure dependent effects.  In the end, they
arrive at ${\cal F}t$ values that are nucleus independent and can
be used to extract $\vud$, modulo uncertainties in the radiative
corrections. The $Z$ independence of their results is an important
consistency check, since the daughter nuclei have  $Z$ values
ranging from 5 to 26 with correspondingly different Coulomb and
structure corrections.

 In Table \ref{tab:nuc}, we give the (slightly) updated values of $\vud$ obtained
from the nine best measured superallowed $\beta$-decays, incorporating the
isospin symmetry-breaking and  structure dependent corrections $\delta_C$ and $\delta_{NS}$
used by Towner and
Hardy,
in conjunction with our new factorization scheme for the radiative
corrections in \eqn{13} (changing $E_m$ and $\delta$ in that expression
as appropriate for each nucleus).  Specifically, we include the corrections $\delta_{NS}$
in the second factor of \eqn{13} and append an additional factor $(1-\delta_C)$.

\begin{table}[hbf]
\caption{
 The values of $\vud$ obtained from superallowed beta-decays.
The error given does not include nuclear and theory uncertainties
common to the analysis. It is taken from
ref.~\protect\cite{Towner:2002rg} and used to obtain the
weighted average.}
\label{tab:nuc}
\begin{tabular}{@{\hspace*{3mm}}l@{\hspace*{20mm}}l@{\hspace*{3mm}}}
\hline\hline
   Nucleus    &          $\vud$ \\
\hline
    $^{10}$C         &  0.97388(76) \\
    $^{14}$O         &  0.97445(41) \\
    $^{26}$Al        &  0.97416(35) \\
    $^{34}$Cl        &  0.97431(40) \\
    $^{38}$K         &  0.97424(43) \\
    $^{42}$Sc        &  0.97351(38) \\
    $^{46}$V         &  0.97372(43) \\
    $^{50}$Mn        &  0.97396(44) \\
    $^{54}$Co        &  0.97409(43)\\
\hline \hline
\end{tabular}
\end{table}
The values of $\vud$ derived from those distinct measurements are very
consistent with one another and range from about 0.9735 to 0.9745.  The
weighted average is centered at $\vud$ = 0.974047; so, we round down to
\ba
           \vud = 0.9740(1)(3)(4),
\label{eq19b}
\ea
where the uncertainties are experimental, nuclear theory and RC
(see eq.~\eq{eq9b}). We have checked that the combined
 small changes in the
RC arising from our new factorization in eq.~\eq{eq13}, improvements in
the higher order leading logs, central value of $m_A$, small NLL
($-0.0001$) correction etc. tend to cancel. For that reason, the result in
\eqn{19b} is essentially the same as the one
obtained earlier by Towner and Hardy (cf.~\eqn{2}) using the radiative corrections
given in Ref.~\cite{SirlinInLang}. We note that the sign of
the ${\alpha\over 2\pi}\delta$ corrections for the superallowed
decays (which are all $e^+$ emitters) is correct in the literature
and numerically the change in sign of that correction for neutron
decay was our biggest modification of previous results.  So,
$\vud$ in eq.~\eq{eq19b} remains the current best value.  Indeed,
comparison with $\vud$ extracted from neutron decay $\tn$ and
$g_A$ values in eq.~\eq{eq18} shows that they share the same RC
uncertainty, but the error on $g_A$ must be improved by about a
factor of 4 before the neutron decay becomes competitive.

One way of reducing the RC uncertainty in $\vud$, perhaps by as
much as a factor of 2 would be to use the $\pi^+ \to \pi^0 e^+
\nu_e$ decay rate for which the loop induced axial-vector
contributions are better controlled and nuclear theory uncertainties
are circumvented \cite{Jaus:1999dw}. However, the small branching
ratio $\sim 10^{-8}$ makes that method statistically challenging.
Nevertheless, an ongoing PSI experiment finds
\ba
\vud = 0.9749(26) \cdot \left[ {\mbox{BR}\left(\pi^+ \to e^+ \nu_e(\gamma)\right)
\over 1.2352\cdot 10^{-4}}\right]^{1/2},
\label{eq20}
\ea
where its dependence on the $\pi^+ \to e^+ \nu_e(\gamma)$ branching ratio
(used for normalization)
is exhibited. We assume the SM theory value of $1.2352\cdot 10^{-4}$ is correct
and hence
obtain $\vud=0.9749(26)$, in excellent agreement with nuclear and neutron
results but with a larger error.  Alternatively, others have chosen to use
the PDG recommended branching ratio of $1.230(4)\cdot 10^{-4}$  which leads to
$\vud=0.9728(30)$ which is also in agreement within errors even though the
central value may appear low \cite{Pocanic:2003pf}.

Note, if we average the $\vud$ determinations above, we still find
\ba
\vud = 0.9740(5), \qquad \mbox{(Average)}
\label{eq21}
\ea
because the superallowed $\beta$-decays dominate and they have an average central value
slightly larger than $0.9740$.

\section{$g_A$: Theory vs. Experiment}
\label{sec4}
Employing the neutron lifetime measurement in eq.~\eq{eq17a} and the value of $\vud$ in eq.~\eq{eq19b},
we find via eq.~\eq{eq19} the Standard Model prediction
\ba
g_A = 1.2703(6)(5),
\label{eq22}
\ea
where the errors are in $\tn$ and $\vud$ (nuclear uncertainty).
That precise prediction is larger than the
PDG recommended value of 1.2670(30) \cite{Hagiwara:2002fs},
but smaller than the single best asymmetry value \cite{Abele:2002wc}
of
\ba
g_A=1.2739(19),
\label{eq23}
\ea
(after small QED corrections are applied).

For now, the $g_A$ in eq.~\eq{eq22} should be considered the standard, to be used
in solar neutrino
flux calculations etc.  However, to take proper advantage of its precision,
one should employ
the neutron decay definition we have assumed and correct the weak interaction
process under
consideration for its own electroweak radiative corrections.
As an example, consider the relationship between the $g_A$
defined via the neutron lifetime whose value is given in \eqn{22} and one
defined by the lowest order polarized neutron decay asymmetry
\ba
      A = {2\g(1-\g)\over 1+3\g^2}, \qquad \g \equiv g_A^{\rm asy}.
\label{eqnew28}
\ea
If we wish to replace $\g$ by $g_A$ in that expression, then there will be
additional (energy dependent) radiative corrections to the asymmetry.
Those corrections were computed long ago, originally by Shann and
later confirmed by many others.  They are quite small, leading to about a
0.1\% shift in the asymmetry \cite{Shann}.  That effect is corrected for in the most
recent \cite{Abele:2002wc}
experimental measurement of the asymmetry, although it is well
below current experimental uncertainties.  It would be useful to compute
the QED corrections to other processes where the very precise value of $g_A$
in \eqn{22} might prove useful, for example primordial nucleosynthesis
or the solar neutrino flux.

\section{$\vus$ Determinations and CKM Unitarity}
\label{sec5}

The 2002 PDG recommended value for $\vus$  \cite{Hagiwara:2002fs},
\ba
\vus_{\rm PDG}=0.2196(26)\qquad \mbox{PDG2002}, \label{eq28}
\ea
has remained
fixed (modulo small variations in its uncertainties) for many
years. It is based on fitted branching ratios from rather old data
for $K_{e3}$ decays, $K^0\to \pi^-e^+ \nu_e$ and  $K^+\to \pi^0e^+
\nu_e$, combined with the kaon lifetimes, $\tau_{K_L}$ and
$\tau_{K^+}$.  The resulting  $K_{e3}$ decay rates are
proportional to $\vus^2$ and can be used for its extraction. In
fact, they are analogous to superallowed nuclear beta decays (or
$\pi^+ \to \pi^0 e^+ \nu_e$) in that only the weak vector current
contributes at the tree level.  Since that current is conserved in the SU(3) flavor
limit, strong interaction corrections are of second order in SU(3)
breaking.  Those effects, characterized by the departure of the
form factor $f_+(0)$ from 1 along with isospin breaking effects were considered
in the classic study by Leutwyler and Roos \cite{Leutwyler:1984je}
that forms the basis for the extracted value of $\vus$ in
eq.~\eq{eq28}.

In addition to SU(3) breaking effects, there is a fairly
significant first order $m_d-m_u$ correction due to
$\pi^0$--$\eta$ mixing ($\sim 4\%$) that must be separately
applied to the charged $K^+_{e3}$ decay rate and an extra
Coulombic $\pi^-e^+$ final state QED interaction for $K^0_{e3}$.
The fact that even with those different isospin violating
corrections, both the neutral and charged $K_{e3}$ decay rates
gave consistent values for $\vus$ (eq.~\eq{eq28} is their average)
has often been used to argue for the validity of eq.~\eq{eq28} as
compared with for example Hyperon beta decays which have tended to
give somewhat larger values for $\vus$ but are not as theoretically clean.

We note that combining \eqn{28} with the value of $\vud$ in
\eqn{19b} and using $\vub^2 \simeq 2.1\times 10^{-5}$ leads to
\ba
\vud^2_{0^+\to 0^+} +\vus^2_{\rm PDG} +\vub^2 = 0.9969\pm 0.0015.
\label{eq29} \ea
That roughly 2 sigma deviation from unitarity has
been a persistent problem for many years.  It has been at times interpreted
as a hint of new physics
or as an indication that something is
wrong with the data and/or theory calculations used to extract
$\vud$ and $\vus$.  Since perfect unitarity would require a rather
large shift (+ 3.2\%) in $\vus$ but a relatively smaller shift (+
0.16\%) in $\vud$, the latter has been generally thought to be the
root of the problem.  As a result, considerable experimental and
theoretical scrutiny has been applied to $\vud$.  Nevertheless,
as emphasized in the first part of this paper,
its value has remained rather stable.

To illustrate the above approach and some of its underlying uncertainties,
we describe the general \ke3 decay rate formula \cite{Leutwyler:1984je},
\ba
\Gamma(K\to \pi e \nu(\gamma)) = {G_\mu^2 m_K^5 \over 192 \pi^3}
S_{\rm EW} \left( 1+\delta^e_K \right) C^2  \vus^2 f_+^2(0) I^e_K,
\label{f30}
\ea
where $C^2=1$ for $K_L$ or $K_S$ decays (to both $\pi^\pm e^\mp$) and
$C^2=1/2$ for $K^\pm$.  $S_{\rm EW}=1.022$ is the universal short-distance
radiative correction in \eqn{13}  \cite{Sirlin82}
while $\delta^e_K$ are model dependent
long-distance QED corrections recently estimated to be (for radiative
inclusive studies) \cite{Andre:2004tk,Cirigliano:2001mk,Cirigliano:2004pv}
\ba
\delta^e_{K^0} &=& +1.3\pm 0.3\%,
\nonumber \\
\delta^e_{K^+} &=& -0.1\pm 0.7\%.
\label{f31}
\ea
The form factor $f_+(0)$ incorporates SU(3) breaking.  Leutwyler and Roos
found
\ba
f_+(0) = f_+^{K^0\pi^-} = 0.961\pm 0.008,
\label{f32}
\ea
a value recently confirmed by a lattice calculation \cite{Becirevic:2004ya} and to some
extent by new papers based on Chiral Perturbation Theory (ChPT)
\cite{Cirigliano:2004pv,Bijnens:2003uy,Bijnens:2004pk}.
In particular, in Ref.~\cite{Bijnens:2003uy} it is shown that the only
unknown constants in the $\order{p^6}$ contributions in ChPT can be determined by
accurate measurements of the slope and curvature of the scalar $K_{l3}$ form
factor. In Ref.~\cite{Jamin:2004re} dispersion relations are employed to
calculate theoretically these two observables, leading to
$f_+(0) = 0.974 \pm 0.0057 \pm 0.0028 \pm 0.009$, which is consistent with Eq.~(\ref{f32}),
although the central value and estimated errors are somewhat larger. In
this paper we employ the classical result given in Eq.~(\ref{f32}), but we also
emphasize the importance of refining the lattice and ChPT calculations of
$f_+(0)$.

In the case of charged kaons, $m_d-m_u$ mass
splittings give rise to $\pi-\eta$ mixing
such that
\ba
f_+^{K^+\pi^0} \simeq 1.022 f_+^{K^0\pi^-} = 0.9821\pm 0.008\pm 0.002.
\label{f33}
\ea
Finally, the phase space factor is determined for a linear form
factor to be (for a slope factor of 0.028)
\ba
I_{K^0}^e &=& 0.1550,
\nonumber \\
I_{K^+}^e &=& 0.1594,
\label{f34}
\ea
while for a quadratic form factor suggested by KTeV data \cite{Alexopoulos:2004sy},
\ba
I_{K^0}^e &=& 0.1535.
\label{f35}
\ea
The difference, $\sim 1\%$, is somewhat accounted for by assigning
an extra $\pm 0.7\%$ form factor decay rate
uncertainty.  However,
we note that the central value of $\vus$ will depend on which parametrization
is used.  The value of $\vus$ in \eqn{28} would become $0.2207$ if a quadratic
parametrization were employed.

Recently, a number of developments have cast doubt on the
reliability of \eqn{28}.  First, a new measurement of the
$K_{e3}^+$ branching ratio by the E865 Collaboration at Brookhaven
National Laboratory finds it to be about 5.3\% larger than the
fitted PDG value used to obtain $\vus$.  Their finding conflicts
with the earlier $K_{e3}$ decay rates, even after all isospin violating
differences are taken into account and on its own leads to \cite{Sher}
\ba
\vus =0.2236(23)/f^{K^+\pi^0}_+(0), \qquad \mbox{E865}
\label{eq30}
\ea
where $f^{K^+\pi^0}_+(0)=1+$SU(3) breaking
effects, including $m_d-m_u$ corrections (but unlike E865, we have not absorbed
QED corrections into its definition). For the value
$f^{K^+\pi^0}_+(0)=0.9842(84) $ effectively used by the E865 collaboration after
removing the QED correction,
one finds $\vus=0.2272(30) $ and correspondingly almost perfect
unitarity, \ba \vud^2_{0^+\to 0^+} +\vus^2_{\rm E865} +\vub^2 =
1.0003\pm 0.0017. \label{eq31}
\ea

However, that is not the end of the story.  In an even more recent
development, the KTeV Collaboration (E832) \cite{Alexopoulos:2004sw}
at Fermilab has reported a
thorough analysis of all primary $K_L$ decay modes, including \ke3.
They find significant disagreement with the PDG fit values for
several of the main $K_L$
branching ratios, including \ke3, exactly the types of shifts required
to bring the $K_L$ system into accord with unitarity and the $K^+$ results of
E865.

     The KTeV Collaboration finds (using the PDG $K_L$ lifetime)
\ba
            \Gamma(K_L\to \pi e\nu) = 0.520(4)\times 10^{-14}\mbox{ MeV}.
\label{eqnew33}
\ea
That radiatively inclusive \ke3 partial decay rate is about 5\% larger than
the 2002 PDG fit value.  After accounting for measured  form factor
effects, a new calculation of QED radiative corrections
\cite{Andre:2004tk,Cirigliano:2001mk,Cirigliano:2004pv}
and SU(3) breaking
(via Leutwyler and Roos \cite{Leutwyler:1984je}), they find
\ba
            \vus = 0.2253(23)  \qquad\mbox{KTeV \ke3},
\label{eqnew34}
\ea
where (because of the very high statistics)
the error is essentially dominated by SU(3) breaking, form factor
shape and $K_L$ lifetime
uncertainties.  Unlike $K^+_{e3}$, the $K_L$ extraction is not directly
sensitive to the
up-down mass difference.  For $K_{\mu 3}$, they obtained a similar result
$\vus=0.2250(23)$, which provides strong confirmation.  Taken together with
the value of $\vud$ in \eqn{19b},  one finds from \eqn{new34}
\ba
   \vud^2+\vus^2+\vub^2 = 0.9994(14) \qquad\mbox{KTeV \ke3 },
\label{eqnew35}
\ea
in very good agreement with unitarity,  a remarkable turn of events.
So, E865 and KTeV are in apparent agreement regarding $\vus$ and unitarity.
Of course, they are
dependent on the $K^+$ and $K_L$ lifetimes which, if the history of the kaon
branching ratios is any indication, could change upon closer scrutiny.

It is useful to average the \ke3 and $K_{\mu 3}$
results from E865 and KTeV in order to
get a single determination of $\vus$ and unitarity constraint that can be
used to limit new physics appendages to the standard model.  To carry
out such an average requires a consistent treatment of features common to
the analyses of both measurements.  In that category we put SU(3) breaking
effects and the form factor shape (linear vs.~quadratic).
We adopt the KTeV quadratic form factor along with its $\pm 0.7\%$ uncertainty
and assume the Leutwyler-Roos estimate of SU(3) breaking and $m_d-m_u$ effects.
Together they shift the E865 value of $\vus$ up by  about 0.7\%.
So, we wind up with the following two quantities to be averaged:
\ba
               \vus &=& 0.2288(26)(20) \qquad \mbox{Shifted E865 $K^+_{e3}$},
\nonumber \\
               \vus &=& 0.2252(13)(20)    \qquad \mbox{KTeV  $K^0_{e3}$ and $K^0_{\mu 3}$ average},
\label{eq2vus}
\ea
where we employ the KTeV \ke3 and $K_{\mu 3}$ averages under the assumption
of muon-electron universality.
The common error from the form factor and SU(3) breaking, $(\pm 20)$,
has been
factorized.  The remaining uncertainty in the $K^+$ case comes from adding in
quadrature the experimental uncertainties along with errors due to QED,
the $K^+$ lifetime, $m_d-m_u$ effects, and normalization uncertainties associated
with correlations among $K^+$ branching ratios.  We estimate the last of
these to be $\pm 0.5\%$.  For the $K_L$,
the first error comes from experimental
uncertainties, the $K_L$ lifetime error (which is appreciable), and QED.

       Carrying out a weighted average, using the first set of errors to
do the weighting, we find
\ba
                 \vus = 0.2259(12)(20)    \qquad \mbox{E865-KTeV Average},
\ea
or, combining errors in quadrature,
\ba
                    \vus = 0.2259(23)  \qquad \mbox{E865-KTeV Average}.
\label{eq44}
\ea
Together with the $\vud$ value in \eqn{2}, that gives
\ba
             \vud^2 + \vus^2 + \vub^2 = 0.9997(10)(10),
\label{eqnew37}
\ea
where the first and second errors arise from $\vud$ and $\vus$, respectively.
Combining the errors in quadrature, we obtain
\ba
             \vud^2 + \vus^2 + \vub^2 = 0.9997(14),
\label{eqnew37x}
\ea
which confirms unitarity superbly.
Agreement with the standard model
expectation can be used to constrain various new physics effects.
For example, if the muon had some additional, exotic decay mode
such as $\mu \to e + $ the wrong neutrinos, it would affect unitarity through
the $G_\mu$ used in our normalization \cite{Marciano:2003ci}.
The good agreement in \eqn{new37x}
 limits the
branching ratios for those types of hypothetical decays to be $<0.2\%$, which
is similar to the best direct constraints, but more general.

      At this point we note that the central values of $\vus$ in \eqn{2vus}
differ by about $1.2 \sigma$.  It could be a simple effect due to experimental
systematics (statistical errors are very small) or might indicate a
problem in the Kaon lifetimes, $m_d-m_u$, QED long distance radiative
corrections or $K^+$ correlated branching ratios.  Improved measurements in
the charged and neutral kaon properties may prove useful in clarifying the
difference.  However, the total uncertainty in $\vus$ of about $\pm 1\%$ is
dominated by SU(3) breaking and the form factor shape and magnitude.  It
will be difficult to reduce those errors much further. Unitarity
prevails, but unless some new procedure for calculating SU(3) breaking effects with
much higher precision is found, it seems unlikely that the error in \eqn{44} can be
reduced much further, i.e. a $\pm 1\%$ uncertainty in $\vus$ is near the end of the
road for \ke3. Similar remarks apply to $\vud$ where theory uncertainty
dominates.  Fortunately, they seem to be ending with a triumph for
unitarity and a strong confirmation of the standard model.

    Other new analyses of \ke3 by the KLOE Collaboration at Frascati
and NA48 at CERN are being completed and should report results shortly.
It will be interesting to see if they confirm E865 and KTeV or reopen the
\ke3 problem.

In another relatively recent development, Cabibbo, Swallow and
Winston (CSW) \cite{Cabibbo:2003ea,Cabibbo:2003cu} have revisited
the extraction of $\vus$ from Hyperon beta decays.  That procedure
is sometimes criticized as unreliable because the decay rates are
renormalized by first order SU(3) breaking effects in the
axial-vector contributions. However, rather than just employing
total decay rates, CSW also included decay asymmetry measurements
that effectively measure the first order SU(3) breaking effects,
analogous to the use of $g_A^{\rm Asy} $ in neutron beta decay to
determine $\vud$.  They found
\ba \vus=0.2250(27)
\qquad\mbox{Hyperon Decays},
\label{eq35}
\ea
where the error quoted is purely experimental and
SU(3) breaking has been neglected.  Nevertheless, if taken at face value
it gives
 \ba \vud^2_{0^+\to 0^+} +\vus^2_{\rm Hyperon} +\vub^2 =
0.9993(16),
\label{eq36}
\ea
i.e.  good agreement with
unitarity.  Because SU(3) breaking effects and various other
theory uncertainties have not been
considered, we do not include \eqn{35} in our averaging.

Finally, a  recent lattice calculation \cite{Aubin:2003ne} of the
pseudoscalar decay constants
\ba
f_K/f_\pi = 1.210(4)_{\rm stat}(13)_{\rm syst}, \label{eq37}
\ea
can be combined \cite{Marciano:2004uf} with the experimental quantity
\ba {\Gamma\left(
K^+ \to \mu^+ \nu_\mu(\gamma)\right) \over\Gamma\left( \pi^+ \to
\mu^+ \nu_\mu(\gamma)\right)}=1.3336(44),\label{eq38}
\ea to yield
\ba
{\vus \over \vud }=0.2278(26),\label{eq39}
\ea
where the uncertainty is lattice dominated.
If unitarity is assumed, that ratio implies
\ba
\vus &=& 0.2221(24),\nonumber\\
\vud &=& 0.9750(5),
\label{eq40}
\ea
while if $\vud=0.9740(5)$ is employed
\ba \vus = 0.2219(25) \qquad \mbox{Lattice}.
\label{eq41}
\ea
It corresponds to a modest 1.4 sigma deviation
from unitarity.  The beauty of the lattice approach is that it is still in its
infancy.
With new larger computers and better treatment of chiral symmetry and
isospin breaking, the uncertainty in \eqn{41} may be reduced by a factor
of 4 or more, making the lattice approach to $f_K/f_\pi$ ultimately the best
way to determine $\vus$.  Should that happen, it will be somewhat ironic
that the axial current determination via $K_{l2}$ and $\pi_{l2}$
decay constants turns out to be
theoretically more pristine that the vector current approach using \ke3.
If something similar were possible for
$\vud$, the unitarity test (constraint) in \eqn{new37x}
might be significantly
improved.

Recently, hadronic $\tau$ decays have been employed to determine $|V_{us}|$.  The value found
in  \cite{Gamiz:2004ar}, using LEP data, is
\ba \vus = 0.2208(34) \qquad \mbox{$\tau$ decays}.
\label{eqTau}
\ea
This determination will likely improve with new data coming from BaBar and Belle.

The $\vus$ central values discussed above vary from about 0.22--0.23
depending on the data used and SU(3) breaking corrections
applied.  That range is to be compared with the value suggested by
$0^+\to 0^+$ beta decays and perfect unitarity \ba \vus =
0.2265(22)\qquad \mbox{Unitarity + $0^+\to 0^+$ Nuclear}.
\label{eq42}
\ea

\begin{figure}[htb]
\hspace*{0mm}
\epsfxsize=85mm
\epsfbox{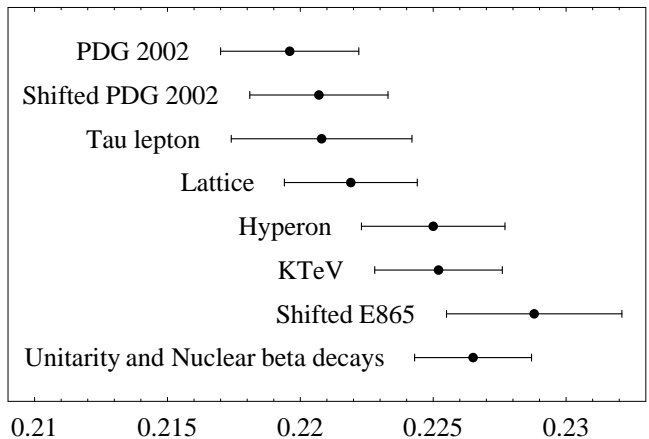}
\caption{Determinations of $\vus$ from various sources.
The Hyperon value does not
include theory errors.  Shifted values correspond mainly to the change of linear
to quadratic form factor parametrization.  All $K_{l3}$ results assume Eqs.~(\ref{f32},\ref{f33}).}
\label{figVus}
\end{figure}

A comparison of the different determinations of $\vus$ is illustrated in
Fig.~\ref{figVus}.

\section{Conclusion and Outlook}
\label{sec6} We have presented an update of the electroweak
radiative corrections to neutron $\beta$-decay.  It currently
provides the most precise determination of $g_A$, a quantity that
finds many applications in nuclear, particle, and astro-physics.
As $g_A$ and $\tau_n$ experimental measurements improve, our
results can be used to obtain a $\vud$ that is competitive with
nuclear $\beta$-decay determinations (which yield $\vud=0.9740(5)$),
but without the nuclear structure dependent uncertainties. In the
end, we will still be limited by a $\pm 0.0004$ theory error that
comes from uncertainties in the axial-vector induced loop
corrections. Reducing the latter error further will require a new
theoretical approach or the measurement of $\Gamma(\pi^+\to
\pi^0e^+\nu_e)$ which has smaller uncertainties.  The $10^{-8}$
branching ratio makes the latter strategy very difficult statistically.

More problematic than $\vud$ in testing CKM unitarity
has been inconsistencies in the various $\vus$ determinations.  Of
particular concern has been the low value $\alt 0.22$ obtained from PDG fits to
old \ke3 data which for a long time has suggested a small but persistent
departure from unitarity.  However,  recently measured \ke3
branching ratios for both the charged and neutral kaon exhibit large
deviations (about 5\%) from the PDG fit values, increasing $\vus$ to a level
consistent with unitarity.  In fact, the average value they provide
$\vus$=0.2259(23) together with $\vud$=0.9740(5) from superallowed beta
decays concur with unitarity expectations up to $\pm 0.0014$ (see \eqn{new37x}).
That good agreement confirms the predicted large +4\% \cite{Sirlin:1974ni} radiative correction
of the Standard Model at the $28\sigma$ level! A triumph for quantum loop effects.
However, before one can be sure that the unitarity problem has been
resolved, the recent results should be confirmed by anticipated new
measurements of the \ke3 branching ratios from the KLOE experiment at Frascati
and the NA48 experiment at CERN.  In addition, given the recent changes in kaon branching
fractions,  it would be nice to have new
confirmation measurements of both the $K^+$ and $K_L$ lifetimes.

The newest approach to $\vus$ appears to be very promising
\cite{Marciano:2004uf}.  It combines a lattice calculation of
$f_K/f_\pi$ with the experimental measurement of $ {\Gamma\left(
K^+ \to \mu^+ \nu_\mu(\gamma)\right) \over\Gamma\left( \pi^+ \to
\mu^+ \nu_\mu(\gamma)\right)}$.  The latter ratio is already very
well measured, but in view of the changes in $K^+$ decay rates,
the numerator should be experimentally revisited. Also, the
electroweak radiative corrections that largely cancel in the ratio
should be reexamined.

The lattice calculation of $f_K/f_\pi$ has some very nice features
that suggest its already small uncertainty may be reduced much
further.  They include a cancellation of the statistical
uncertainties which are highly correlated and the scale
uncertainty.  The challenge is to refine extrapolations to the
chiral and continuum limits with full dynamical fermion
simulations.  That should become possible with the advent of very
large special purpose computing facilities.

   The current confirmation of unitarity embodied in \eqn{new37x} gives us further
confidence in the Standard Model and can be used to constrain ``New
Physics'' appendages to it.  Roughly speaking, it rules out additional tree
or loop level contributions to the $\vud^2 +\vus^2$ combination at about
the $\pm 0.2\%$ level. Effects of about that magnitude could in principle come
about from supersymmetry loops \cite{Barbieri:1992qp,Hagiwara:1995fx,Kurylov:2001zx},
an extra $Z'$ boson \cite{Marciano:1987ja}, heavy quark or lepton mixing, exotic muon decays
(since we normalize in terms of $G_\mu$), compositeness, extra dimensions
\cite{Marciano:1999ih} etc.
However, there are no really compelling reasons to expect such a large
deviation.  For that reason, confirmation of unitarity via experimental or
theoretical changes has been anticipated.  Its verification
is, nevertheless, important. Also, finding new ways to improve the
unitarity constraint, e.g. lattice calculations, is difficult, but should
be strongly encouraged. Such challenges push our computational and
experimental skills to new limits, stimulate ingenuity,
and perhaps at some point new physics will be uncovered.

{\em Note added:} After the completion of this paper, a new low temperature measurement
of the neutron lifetime was reported, $\tau_n = 878.5(7)(3) $ sec \cite{Serebrov:2004zf}.
It differs from the world
average in Eq.~(\ref{eq17a})
by about $6.7 \sigma$.  In conjunction with $|V_{ud}|=0.9740(5)$, it would lead on
its own to a larger $g_A$ value, namely $g_A=1.2766(6)(5)$; instead, if used in combination
with the best asymmetry value
$g_A = 1.2739(19)$, it would imply $|V_{ud}|=0.9757(4)(11)(4)$.  Although this value is not in sharp
disagreement with Eq.~(\ref{eq2})
(the difference is about $1.3 \sigma$), clarification of the neutron lifetime
differences by new improved experiments is clearly an important goal for the future.

Also, the KLOE collaboration has
presented some preliminary new results on $K_{e3}$ and $K_{\mu 3}$ decay rates and the
$K_L$ lifetime \cite{Franzini:2004kb}. They confirm the branching ratio increases observed by KTeV
(within errors) and when finalized should improve the $K_L$ lifetime world
average.

  Finally, a recent analysis \cite{2004na48} by the NA48 Collaboration at CERN finds
a neutral $K_{e3}$ decay rate consistent with the KTeV result in eq.~(\ref{eqnew33}).
However, they effectively employ (after extracting electromagnetic
contributions) a larger value of $f_+(0)=0.974$ suggested by recent chiral perturbation
results
\cite{Cirigliano:2001mk,Cirigliano:2004pv,Becirevic:2004ya,Bijnens:2003uy,Bijnens:2004pk,Jamin:2004re}
and a linear form factor parametrization of phase space. As
a result, they obtain $\vus = 0.2187(28)$ which suggests an overall $2.2 \sigma$
deviation from unitarity.  Such a significant change in the interpretation
reinforces the importance of refining lattice and chiral perturbation
theory calculations of $f_+(0)$ as well as the need to better experimentally
determine its $q^2$ dependence.

\appendix
\section*{Appendix}
\label{secA}
In this appendix, we describe the input that went
into:
\begin{enumerate}
\item
the leading log summations $L(2E_m,m_p)$ and $S(m_p,m_Z)$;
\item
the partial next to leading log calculation due to photonic
vacuum polarization insertions and
\item
the residual ${\alpha\over 2\pi}
\delta$ correction that results from a proper matching of the
Fermi function and long distance $\order{\alpha}$ corrections.
\end{enumerate}

\subsection{Leading Log Summations}
\label{secA1}
We follow the approach of ref.~\cite{Marciano:1986pd} where a
renormalization group summation for the leading short-distance
logs was given and a value for $S(m_p, m_Z)$ derived.  Here we
extend that method down to the intermediate region $2E_m-m_p$.
That simply requires a change in the anomalous dimension from
$2\alpha/\pi$ to $3\alpha/2\pi$ and an evaluation of the
$\overline{\rm MS}$ (modified minimal subtraction) coupling
$\alpha(\mu)$ at low scales.

The leading log summation in an $\overline{\rm MS}$ approach is simply
given by
\ba
L(2E_m,m_p) &=&
\left( {\alpha(m_u)\over \alpha(2E_m)} \right) ^{9/4}
\left( {\alpha(m_d)\over \alpha(m_u)} \right) ^{27/28}
\left( {\alpha(m_\mu)\over \alpha(m_d)} \right) ^{27/32}
\left( {\alpha(m_s)\over \alpha(m_\mu)} \right) ^{27/44}
\left( {\alpha(m_p)\over \alpha(m_s)} \right) ^{9/16},
\label{eqA1} \\
S(m_p,m_Z) &=&
\left( {\alpha(m_c)\over \alpha(m_p)} \right) ^{3/4}
\left( {\alpha(m_\tau)\over \alpha(m_c)} \right) ^{9/16}
\left( {\alpha(m_b)\over \alpha(m_\tau)} \right) ^{9/19}
\left( {\alpha(m_W)\over \alpha(m_b)} \right) ^{9/20}
\left( {\alpha(m_Z)\over \alpha(m_W)} \right) ^{36/17},
\label{eqA2}
\ea
where the $\alpha(\mu)$ are values of the $\overline{\rm MS}$
(modified minimal subtraction) QED coupling
at a scale $\mu$.  This coupling is given to leading log order by
\ba
\alpha^{-1}(\mu) &=& \alpha^{-1}(m_e) - {2\over 3\pi} \sum_f Q_f^2
\Theta(\mu-m_f) \ln {\mu\over m_f}
+ {7\over 2\pi} \Theta(\mu-m_W) \ln {\mu\over m_W},
\label{eqA3}\\
\alpha^{-1}(m_e) &=& 137.036 + {1\over 6\pi} = 137.089.
\label{eqA4}
\ea
In that expression, the sum is over all quarks and lepton flavors $f$ (with a color
factor of 3 for quarks). We thereby find
\ba
\alpha^{-1} (2E_m = 2.585\; \mbox{MeV}) &=& 136.745
,  \nonumber \\
\alpha^{-1} (m_u = 62\; \mbox{MeV}) &=& 136.0708
,  \nonumber \\
\alpha^{-1} (m_d = 83\; \mbox{MeV}) &=& 135.9263
,  \nonumber \\
\alpha^{-1} (m_\mu) &=& 135.7896
,  \nonumber \\
\alpha^{-1} (m_s = 215\; \mbox{MeV}) &=& 135.2368
,  \nonumber \\
\alpha^{-1} (m_p) &=& 133.9861
,  \nonumber \\
\alpha^{-1} (m_c = 1.35\; \mbox{GeV}) &=& 133.67728
,  \nonumber \\
\alpha^{-1} (m_\tau) &=& 133.3662
,  \nonumber \\
\alpha^{-1} (m_b = 4.5\; \mbox{GeV}) &=& 132.1174
,  \nonumber \\
\alpha^{-1} (m_W = 80.4\; \mbox{GeV}) &=& 128.0389
,  \nonumber \\
\alpha^{-1} (m_Z = 91.1875\; \mbox{GeV}) &=& 128.001.
\label{eqA5}
\ea
Relatively low effective quark masses \cite{Marciano:1993jd}
have been used in those results in order to incorporate QCD
contributions
at low energies.  In that way $\alpha^{-1} (m_Z ) = 128.0$
is obtained while a more detailed higher
order QED and QCD analysis including $e^+e^- \to $ hadrons
via a dispersion relation gives
$\alpha^{-1} (m_Z ) = 127.934$.

Employing the above values in \eqn{A1} and \eqn{A2}, we find
\ba
L(2E_m,m_p) &=& 1.02094,\nonumber \\
S(m_p, m_Z) &=& 1.02248. \label{eqA6}
\ea
Those results are not very sensitive to the quark masses employed.
 For example, changing the $m_u$
value by a factor of 2 leads to a shift in the RC by less than $3\times 10^{-5}$ which is well
below the overall $\pm 8\times 10^{-4} $ uncertainty assumed in the RC.

A useful relation, valid for the nine transitions in Table \ref{tab:nuc},is
\ba
L(2E_m,m_p) = L(m_e,m_p) \left({\alpha(m_e)\over \alpha(2E_m)}\right)^{9/4},
\ea
which leads to
\ba
L(2E_m,m_p) = 1.026725 \left(1 - {2\alpha(m_e)\over 3\pi} \ln{2E_m\over m_e}\right)^{9/4}.
\ea

\subsection{Next-to-leading logarithmic corrections due to photonic
vacuum polarization.}

\begin{figure}[htb]
\hspace*{0mm}
\begin{tabular}{c@{\hspace*{15mm}}c@{\hspace*{15mm}}c}
\epsfxsize=30mm \epsfbox{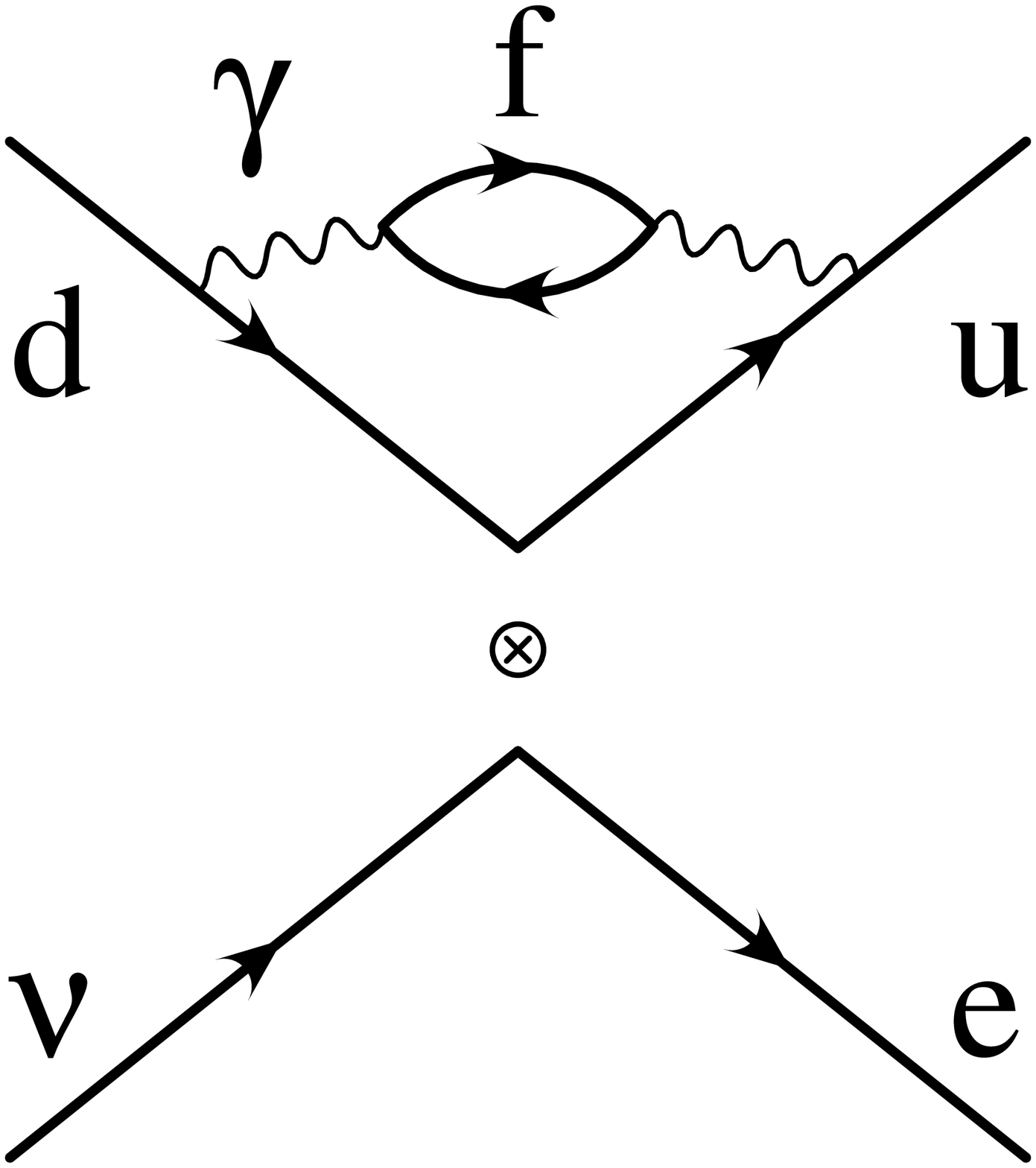} &
\epsfxsize=30mm \epsfbox{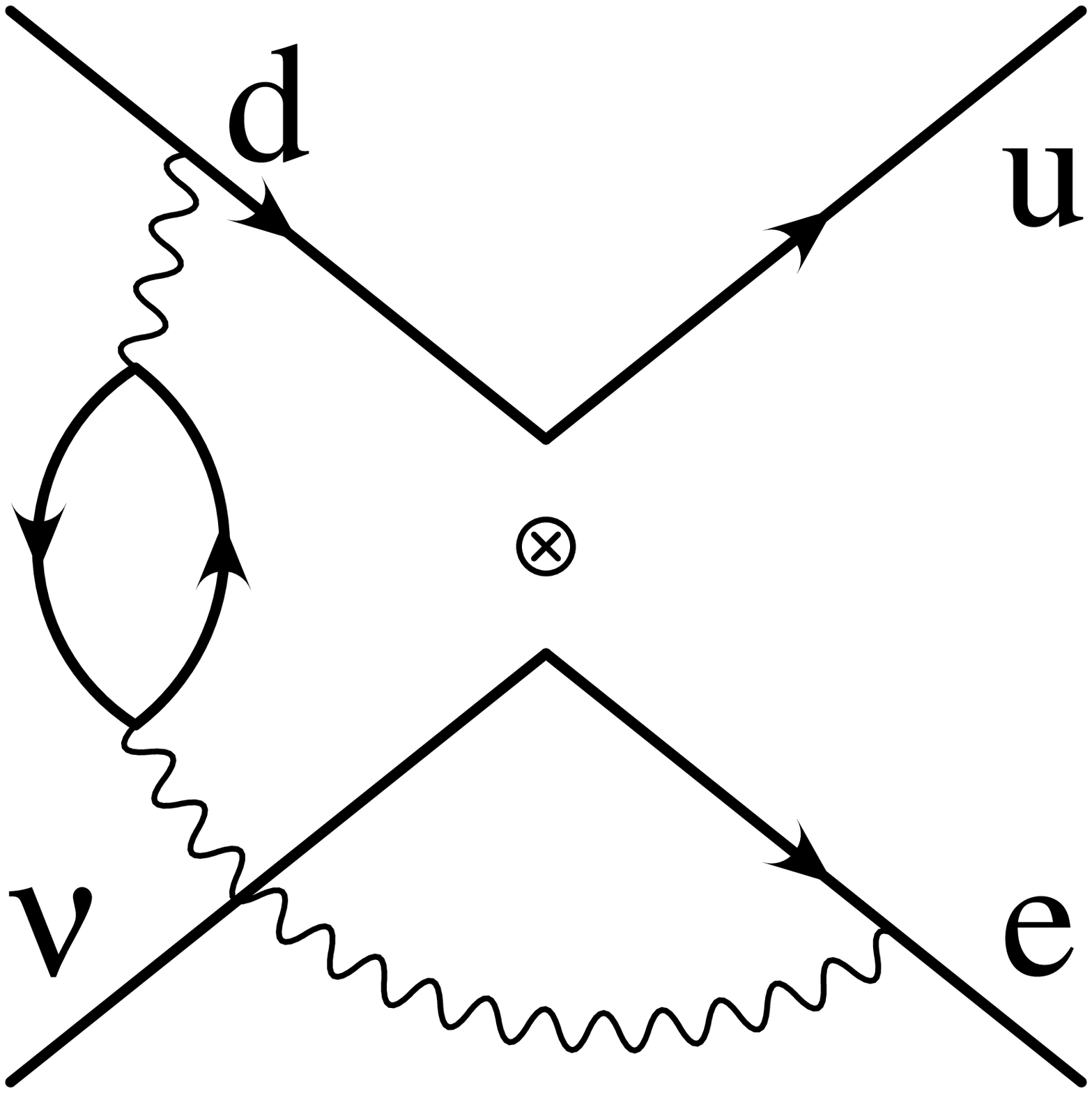} &
\epsfxsize=30mm \epsfbox{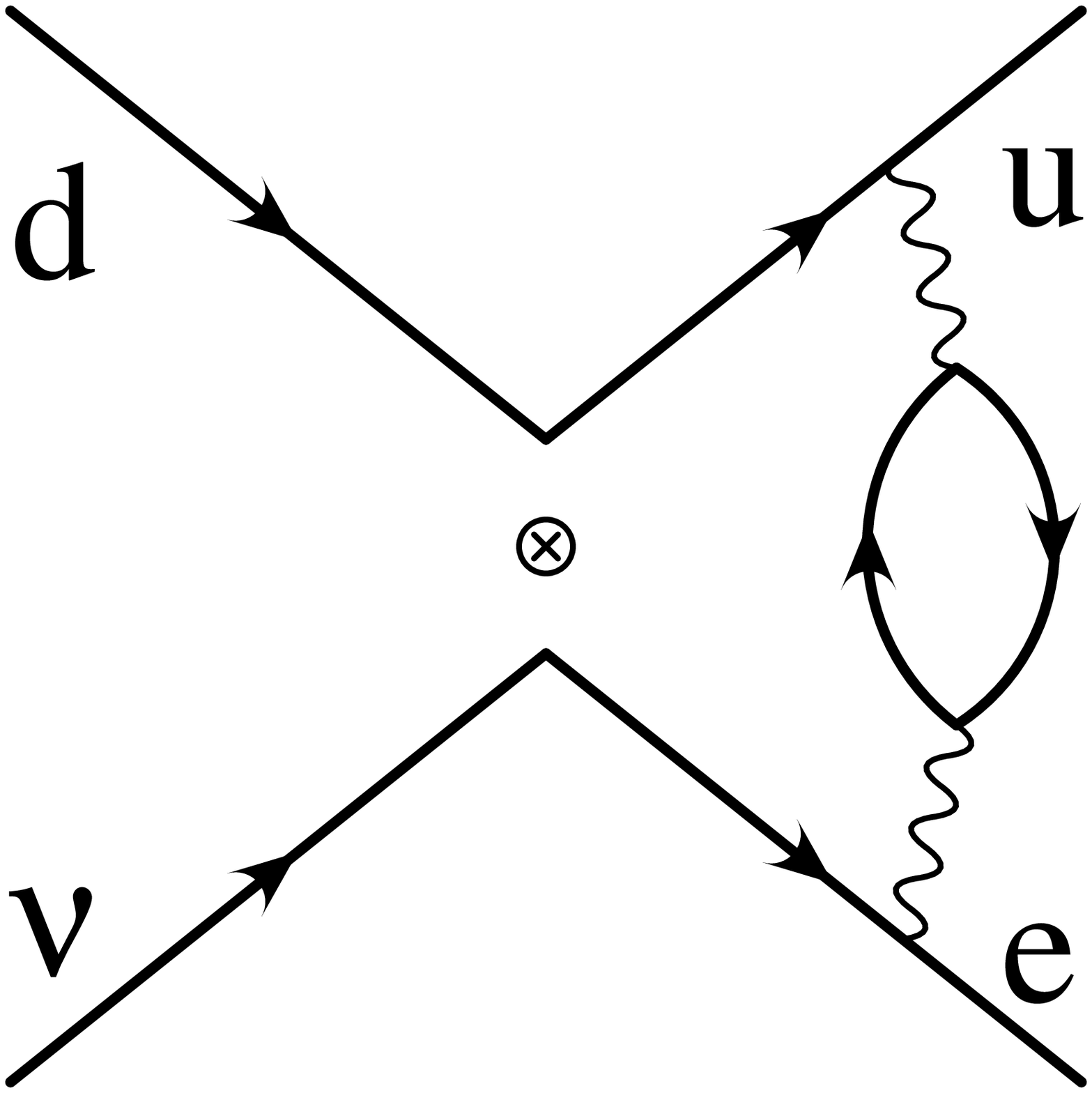}
\end{tabular}
\caption{Vacuum polarization diagrams contributing logarithmic and NLL
corrections to the beta decay rate. QED loop corrections to external charged particle
legs are included but not
illustrated.} \label{BWvacpol}
\end{figure}

Next-to-leading logarithmic (NLL) corrections to weak decays have
been studied in detail for QCD (see, for example,
\cite{Altarelli:1981fi,Buras90}). Here we adapt those results to
QED. We consider a subset of contributions, containing fermion
loop insertions in the photon propagator, as depicted in
Fig.~\ref{BWvacpol}.  That particular subset is somewhat enhanced
by the large number of contributing fermions.  Therefore, we expect it to dominate the NLL
contributions.

The effect of short distance corrections can be parameterized as follows.  We keep only
leading-logarithmic terms, and those NLL ones that arise from fermion loops. Then, the
decay width $\Gamma$ can be expressed in terms of the lowest-order width $\Gamma^0$ times
a correction factor,
\ba
\Gamma &=& \Gamma^0 \left[ 1 - \gamma^{(0)} L a
 -L a^2 \left( \gamma^{(1)} + 2\beta_0 B \right)
 +{1\over 2}L^2 a^2 \gamma^{(0)}\left( \beta_0 + \gamma^{(0)}\right)\right],
\nonumber \\
 L & \equiv & \ln{M_W^2 \over \mu^2}, \qquad  a \equiv  {\alpha(\mu)\over 4\pi},
\label{QEDres}
\ea
and we will be interested in the value of this correction at $\mu \simeq m_p$.  A similar
analysis can be extended down to $2E_m$.

The leading order anomalous dimension of the four-fermion operator
$\bar u \gamma^\mu (1-\gamma_5) d \otimes \bar e \gamma_\mu (1-\gamma_5) \nu$
is given by diagrams similar to those in Fig.~\ref{BWvacpol}, but without fermion loops.
 It can be expressed in terms of  fermion charges,
\ba
\gamma^{(0)} = (Q_u-Q_d)^2 +Q_e(Q_e+8Q_u-2Q_d) = -4.
\ea
The running of the coupling constant is described by a sum over all contributing fermions
with charges $Q_f$ and number of color varieties $n_f$
 (equal 3 for quarks and 1 for others),
\ba
\beta_0 = -{4\over 3}\widetilde n,   \qquad \widetilde n \equiv \sum_f n_f Q_f^2.
\ea
At the NLL order, we also need
the two-loop anomalous dimension $\gamma^{(1)}$ and the matching coefficient $B$.
The former is determined by calculating $1/\epsilon$ poles of the diagrams in
Fig.~\ref{BWvacpol}.  The latter is given by the finite part of analogous diagrams without
fermion loop insertions.
Their individual values depend on the scheme of the calculation (for
example, treatment of $\gamma_5$), and only their combination
occurring in eq.~(\ref{QEDres}) is scheme-independent.

We list here the values of both quantities obtained in the naive dimensional
regularization (NDR), with an anticommuting $\gamma_5$, and in the 't Hooft-Veltman
scheme (HV),
\ba
\begin{array}{l@{\hspace*{15mm}}l}
\gamma^{(1)}_{\rm HV} = -{20\over 3} \widetilde n, &
\gamma^{(1)}_{\rm NDR} = -{44\over 9} \widetilde n,
\\[2mm]
B_{\rm HV} = -{23\over 6}, &
B_{\rm NDR} = -{19\over 6}.
\end{array}
\ea
In both schemes we get
\ba
2\beta_0 B + \gamma^{(1)} = {32 \over 9} \widetilde n.
\ea
With these results, the logarithmic corrections of eq.~(\ref{QEDres}) become
\ba
\Gamma
&=&
\Gamma^0 \left\{ 1+{2\alpha(\mu) \over \pi}\ln{M_W\over \mu}
+\left({\alpha\over \pi}\right)^2
\ln{M_W\over \mu}
\left[ \left( {2\widetilde n \over 3}+2\right)  \ln{M_W\over \mu}
 -  {4\widetilde n\over 9}\right]
 \right\}.
 \ea
All but the last terms in this correction factor are included and summed up to all orders
in the factor $S$ in \eqn{A2}.  The last term is the fermionic NLL correction.
Because it is very small, we estimate it as though all fermions $u,d,s,c,b$ and $e,\mu,\tau$
 contributed over the whole range from $M_W$ to $m_p$.  In this approximation, we have
\ba
\widetilde n &=& {20\over 3},
\ea
and the numerical value of the NLL fermionic correction is
\ba
-  {4\widetilde n\over 9}\left({\alpha\over \pi}\right)^2
\ln{M_W\over m_p} = -0.00007.
\ea
Small additional NLL contributions from the $E_m$
to $m_p$ region shift that value to about $-0.0001$.
Other two-loop NLL and non-logarithmically enhanced corrections are
individually of order $\left({\alpha\over \pi}\right)^2\ln{m_p\over 2E_m}\simeq 0.00003$
or smaller.  Hence, we assume that they can be neglected.

\subsection{Evaluation of the $\order{Z\alpha^2}$ corrections}

In this Appendix we summarize the evaluation of the
$\order{Z\alpha^2}$ corrections in the case of neutron
$\beta$-decay ($Z=1$).  These are defined as the residual
corrections of this order not contained in the product of the
Fermi function and the $\order{\alpha}$ corrections. For a point
nucleus, analytic results for the correction to the charged lepton
spectrum has been obtained in the extreme relativistic
approximation \cite{Sirlin:1986cc}.  The sign is opposite for
electron and positron emitters (as can be readily understood by a
glance at the figures in Ref. \cite{Sirlin:1986cc}) and, for the
neutron decay, we have
\ba
\Delta P = -\alpha^2 \left[ \ln \left(
{m_p \over m_e}\right) + {43\over 18} - {5\over 3} \ln\left(
{2E\over m_e}\right) \right],
\label{A1}
\ea
(see Eq. (5) of Ref.
\cite{Sirlin:1986cc}).  It has also been verified that, to good
approximation, the leading contributions to Eq.~\eq{A1}
extrapolate smoothly to their non-relativistic limit,  which has
also been obtained analytically.  Next, we evaluate numerically
the spectral average of $\ln(2W)$ $(W\equiv E/m)$ using the phase
space factor $\sqrt{W^2-1}\;W\left( W_{\rm max} - W\right)^2$,
leading to
\ba
\langle \ln(2W)\rangle =1.1390.
\label{A2}
\ea
Combining Eqs.~\eq{A1} and \eq{A2}, we obtain
\ba
\langle \Delta
P\rangle =-8.006 \alpha^2  = -4.26\times 10^{-4}.
\label{A3}
\ea

In order to take into account the finite proton size, we employ
Eqs.~(7-9) of Ref.~\cite{Sirlin:1987sy}, with the sign again
reversed, since we are dealing with an electron emitter.  These
formulae correspond to a uniformly charged sphere of radius
$R=\sqrt{5\over 3}a$, where $a\equiv \sqrt{6}/\Lambda$ is the rms
charge radius of the proton.  Using $a=0.90$ fm
\cite{Wilkinson:1982hu}, we have $\Lambda / m_p = 0.572$.  Insertion
of this value in Eqs.~(7-9) of Ref.~\cite{Sirlin:1987sy}, leads to
a finite proton size effect $-5.4 \times 10^{-6}$.  Combining this
result with Eq.~\eq{A3}, we obtain our final answer for this class
of corrections in the case of neutron decay:
\ba
{\alpha\over 2\pi}\delta
= -4.3 \times 10^{-4}. \label{A4}
\ea
We have verified
that terms of $\order{\left( \Lambda/m_p\right)^3 } $, not
contained in Eqs.~(7-9) of Ref.~\cite{Sirlin:1987sy}, as well as
estimates of the corrections $\order{Z^2\alpha^3}$, give
negligible contributions in the case of neutron decay.

\begin{acknowledgments}

We thank Kim Maltman for bringing Ref.~\cite{Serebrov:2004zf} to our attention,
Albert Young for discussions regarding the
neutron lifetime, and Paolo Franzini for updating us on preliminary KLOE
results.
Part of this work was carried out while the authors were
visiting the Kavli Institute for Theoretical Physics in Santa Barbara,
with partial support by the National Science Foundation under Grant No. PHY99-07949.

A.C. was supported by the Natural Sciences and
Engineering Research Council of Canada.

W.J.M. acknowledges support by DOE grant
DE-AC02-76CH00016 and thanks
the Alexander von Humboldt Foundation for support
during the completion of this work.

A.S. is grateful to the Theory Group of the $2^{\rm nd}$
Institute for Theoretical
Physics for the hospitality extended to him during a visit when this
manuscript was prepared.

The work of A.S. was supported in part by the Alexander von Humboldt
Foundation Research Award No. IV USA 1051120 USS, and the
National Science Foundation Grant PHY-0245068.

\end{acknowledgments}


\begin{thebibliography}{10}
\bibitem{Shann}
R.~T. Shann, Nuovo Cimento {\bf A5},  591  (1971), {If} $g_A$ is obtained from
  the total asymmetry in neutron decay, the effect of the radiative corrections
  derived in this paper is to increase $g_A$ by $4\times 10^{-4}$, which is
  much smaller than the current experimental errors.
\end{thebibliography}

\end{document}